\documentclass[fleqn,usenatbib]{mnras}

\usepackage{newtxtext,newtxmath}

\usepackage[T1]{fontenc}

\DeclareRobustCommand{\VAN}[3]{#2}
\let\VANthebibliography\thebibliography
\def\thebibliography{\DeclareRobustCommand{\VAN}[3]{##3}\VANthebibliography}

\newcommand{\be}{\begin{equation}}
\newcommand{\ee}{\end{equation}}

\newcommand{\Msun}{M_\odot}
\newcommand{\Mdot}{\dot{M}}
\newcommand{\Mdotstar}{\dot{M}_\ast}
\newcommand{\Mdotin}{\dot{M}_\mathrm{in}}

\newcommand{\Edot}{\dot{E}}
\newcommand{\Pdot}{\dot{P}}

\newcommand{\rout}{r_\mathrm{out}}
\newcommand{\rin}{r_\mathrm{in}}
\newcommand{\rlc}{r_\mathrm{LC}}
\newcommand{\rco}{r_\mathrm{co}}

\newcommand{\rA}{r_\mathrm{A}}

\newcommand{\Ostar}{\Omega_\ast}

\newcommand{\Lacc}{L_\mathrm{acc}}

\newcommand{\Lcool}{L_\mathrm{cool}}
\newcommand{\Md}{M_\mathrm{d}}

\newcommand{\Lx}{L_\mathrm{X}}

\newcommand{\Tp}{T_\mathrm{p}}

\newcommand{\Gammaacc}{\Gamma_\mathrm{acc}}
\newcommand{\GammaD}{\Gamma_\mathrm{D}}
\newcommand{\Gammadip}{\Gamma_\mathrm{dip}}

\newcommand{\gpers}{g~s$^{-1}$}
\newcommand{\ergpers}{erg~s$^{-1}$}

\newcommand{\spers}{s~s$^{-1}$}

\newcommand{\Alfven}{Alfv$\acute{\mathrm{e}}$n~}
\newcommand{\srca}{GLEAM-X J162759.5--523504.3}

\usepackage{graphicx}	
\usepackage{amsmath}	
\usepackage{ulem}

\title[Evolution of \srca]{Evolution of the long-period pulsar \srca}

\author[Gen\c{c}ali et al.]{
A. A. Gen\c{c}ali,$^{1}$\thanks{E-mail: gencali@sabanciuniv.edu}
\"{U}. Ertan,$^{1}$
M. A. Alpar$^{1}$
\\
$^{1}$Sabanc{\i} University, Orhanl{\i}, Tuzla, 34956, \.{I}stanbul, Turkey
}

\date{Accepted XXX. Received YYY; in original form ZZZ}

\pubyear{2022}

\begin{document}
\label{firstpage}
\pagerange{\pageref{firstpage}--\pageref{lastpage}}
\maketitle

\begin{abstract}
The long-period ($P = 1091$~s) of the recently discovered pulsar \srca~can be attained by neutron stars evolving with fallback discs and magnetic dipole moments of a few $10^{30}$~G~cm$^3$ at ages greater than $\sim 2 \times 10^5$~yr consistently with the observational upper limits to the period derivative, $\Pdot$, and the X-ray luminosity, $\Lx$, of the source. The current upper limits for $\Pdot$ allow two alternative present states: 
(1) The disc is still active with ongoing accretion at a low rate such that the accretion luminosity is much less than the neutron star's cooling luminosity, which in turn is below the upper limit for $\Lx$. In this scenario the spin-down will continue at $\Pdot \sim 10^{-10}$~\spers~until the disc becomes inactive; the final period will be $P \sim$ a few $10^3$~s.
(2) The disc is already inactive, there is no accretion. In this case the period evolution has leveled off to the observed value in the final period range. The remaining, very weak, dipole torque sustaining asymptotic spin-down at $\Pdot \sim 4 \times  10^{-18}$~\spers. Long periods $P \sim$ a few $10^3$~s were predicted for the final states of soft gamma repeaters and anomalous X-ray pulsars with relatively strong dipole fields in earlier work with the fallback disc model. 
\end{abstract}



\section{Introduction}
Rotational periods of radio pulsars and single neutron stars from other populations, namely anomalous X-ray pulsars and soft gamma repeaters (AXP/SGRs), dim isolated neutron stars (XDINs), high-magnetic-field radio pulsars (HBRPs), compact central sources (CCOs), and rotating radio transients (RRATs) are smaller than $\sim 20$~s, while some populations like AXP/SGRs and XDINs have period distributions in the $1 - 20$~s range \citep[for reviews of isolated neutron star populations, see][]{Harding2013, Kaspi2016, Kaspi2017}. The recent discovery of the very long-period pulsar, \srca, with spin period $P \simeq 1091$~s \citep{Hurley-Walker2022} poses crucial questions for present theoretical models: What is the evolutionary history of this source? Is it a member of an unidentified new population following an independent evolutionary path, or does it have an evolutionary connection with already known populations? 

\srca~shows transient radio bursts with durations of about one month, while the pulse widths vary in the $30 - 60$~s range. The period derivative was estimated to be $0 < \Pdot < 1.2 \times 10^{-9}$~\spers, corresponding to a rotational power $\Edot~=~4~\upi^2~I~\Pdot/~P^3~<~1.2~\times~10^{28}$~\ergpers, much smaller than the  luminosity of the pulsed radio emission $\sim 4 \times 10^{31}$~\ergpers~for a distance of $1.3 \pm 0.5$~kpc \citep{Hurley-Walker2022}. The observed radio pulses therefore cannot be produced by  magnetic dipole radiation. The upper limit to the X-ray luminosity, $\Lx < 10^{32}$~\ergpers, implies that the age of the source is greater than $\sim 10^{5}$~yr according to theoretical neutron star cooling curves \citep[e.g.][]{Page2006, Page2009, Potekhin2018, Potekhin2020}.  

Our conclusions in this work are based on the assumption that the long-P pulsar is a neutron star, and the observed period is the spin period of the source.  High luminosity,  high linear and constant polarization ($88 \pm 1 \%$) of the radio pulses of the pulsar \citep{Hurley-Walker2022} require highly ordered strong magnetic fields, which cannot  be produced by white dwarfs in binaries or by exoplanets. Considering also the small upper limit on the size of the emitting region ($0.5$ light second),  \citet{Hurley-Walker2022} argued that the long-P pulsar is likely to be a neutron star producing the observed periodicity through rotation rather than orbital motion.

Pulsed radio emission of a neutron star produced  by its rotating dipole field is likely to be hindered when there is mass flow on to the star. Indeed, transitional millisecond pulsars show only X-ray pulses when accretion is allowed in their LMXB states, while only radio pulses are observed when accretion and X-ray pulses are switched off, in their radio pulsar states. In the same fallback disc framework employed here, the properties of HBRPs are always produced in the strong propeller (SP) phase during which accretion is not allowed. The pulsed radio emission of the long-P pulsar, however, is obviously not powered by a normal rotating dipole-field  mechanism given that the pulsed radio luminosity is well above the rotational power of the source.

There are four known AXP/SGRs (magnetars) displaying radio pulses exclusively in transient epochs after the onset of their  X-ray outburst states with spectral properties rather different from those of normal radio pulsars \citep{Kaspi2017}. There is no normal pulsed radio emission from any other AXP/SGRs. Considering the pulse properties and the transient nature of the radio pulses, \citet{Hurley-Walker2022} suggested that the long-P pulsar could be a magnetar producing its radio pulses with a mechanism similar to that operating in AXP/SGRs.

We show that the long period of \srca~can be reached by a neutron star evolving with a fallback disc and with a magnetic dipole field strength of a few $10^{12}$~G at the equator while simultaneously satisfying the upper bounds on $\Lx$ and $\Pdot$. Such long period sources were predicted by \citet{Benli2016, Gencali2021} as the consequence of neutron star evolution with fallback discs.

The possibility of disc formation by fallback supernova material around new-born neutron stars was proposed by 
\citet{Colgate1971, Michel1988, Chevalier1989}. The evolution of young neutron stars with fallback discs was invoked by \citet{Chatterjee2000} to explain the period clustering and X-ray luminosities of AXPs. \citet{Alpar2001} proposed that initial conditions of newly formed neutron stars, including the properties of possible fallback discs, together with initial spin periods 
and dipole moments, could yield different evolutionary paths leading not only to AXPs but also to other classes of young neutron stars. Emission from fallback discs was studied extensively with applications to different sources 
\citep{Perna2000, Ertan2006, Ertan2007,Ertan2017, Ertan+2017, Posselt2018}. The broad-band, optical to mid-IR, spectrum of 
4U 0142+61 \citep{Hulleman2000, Hulleman2004, Morii2005, Wang2006} can be explained by the emission from an irradiated disc \citep{Ertan2007}. The soft and hard X-ray spectra and pulse profiles of AXPs can be accounted for in terms of emission by channeled mass flow through the accretion column on to the neutron star \citep{Trumper2010, Trumper2013, Kylafis2014}.           

The long-term evolution model including effects of X-ray irradiation, with the contribution of the cooling luminosity, and 
eventual disc inactivation was employed to explain the evolution of AXP/SGRs \citep{Ertan2009, Benli2015, Benli2016}, XDINs \citep{Ertan2014}, HBRPs \citep{Caliskan2013, Benli2017, Benli2018}, CCOs \citep{BenliCCO2018}, and RRATs \citep{Gencali2018, Gencali2021}.We use the model to investigate the long-term evolution of \srca. The detailed description is given in the references above. The model is briefly described in Section \ref{model}. Results are discussed in Section \ref{results}, and conclusions are given in Section \ref{conc}. 

\section{Model}
\label{model}

The initial period, $P_0$, the initial mass of the disc, $\Md$, and the strength of the magnetic dipole field (on the pole), $B_0$, constitute the initial conditions of our model. In addition, there are three main disc parameters, namely the irradiation efficiency parameter, $C$, the critical temperature, $\Tp$, below which the disc becomes inactive, and the kinematic viscosity parameter, $\alpha$ \citep{Shakura1973}. These disc parameters are likely to be similar for fallback discs of different systems in corresponding regimes. The irradiation efficiency depends on disc geometry and on the mass-flow rate of the disc. It could vary with significant changes in the accretion rate. All these parameters are kept constant in our simulations. For different neutron star families we obtain reasonable results with $C \simeq (1-7) \times 10^{-4}$, $\alpha \simeq 0.045$, and $\Tp \simeq 50-100$~K. 

The Keplerian angular frequency of the disc matter equals $\Ostar$ at the co-rotation radius, $\rco = (G M / \Ostar^2)^{1/3}$, where $G$ is the gravitational constant, $M$ and $\Ostar$ are the mass and the angular speed of the star. When the \Alfven radius, 
$\rA \simeq \big[ \umu^4/ (G M \Mdotin^{-2}) \big]^{1/7}$, where $\umu$ is the magnetic dipole moment and  $\Mdotin$ is the mass-inflow rate of the disc, is between $\rco$ and the light cylinder radius $\rlc = c / \Ostar$, the system is in the accretion with spin-down (ASD) regime. In this phase, we take the inner radius of the disc $\rin = \rco$, and calculate the magnetic spin-down torque, $\GammaD$, produced by the disc-field interaction by integrating the magnetic torques from $\rco$ to $\rA$. which can be written in terms of the mass accretion rate, $\Mdot_*$, as
\begin{equation}
\GammaD = \Mdotstar~ (G M \rA)^{1/2} 
\left[1- \left(\frac{\rA}{\rco}\right)^3\right]. 
\end{equation}
The spin-up torque associated with the accretion on to the star from $\rin = \rco$ is  
\begin{equation}
\Gammaacc = \Mdotstar~ (G M \rco)^{1/2} 
\end{equation}
\citep{ErtanE2008}. 
In the ASD phase, we take $\Mdotstar = \Mdotin$. The total torque acting on the star is $\Gamma = \GammaD + \Gammaacc + \Gammadip$. The disc torque $\GammaD$ dominates over $\Gammaacc$ when $\rA$ is not close to $\rco$, while the magnetic dipole torque, $\Gammadip$, is negligible in most cases.   

With decreasing $\Mdotin$, $\rA$ becomes greater than $\rlc$, while $\rin \simeq \rco$. When $\rA > \rlc$, we replace $\rA$ in equation (1) with $\rlc$. In a simplified model, we take $\rA = \rlc$ as the critical condition for the onset of the strong propeller (SP) phase. With increasing $\Mdotin$ the transition from the ASD phase to the spin-up phase takes place when 
$\rA \simeq \rco$. In our calculations, with few exceptions, this transition is not encountered during the evolution of neutron stars with fallback discs.A more realistic torque model including all the rotational phases and transition conditions between the phases was developed recently, with applications to different neutron star systems in low mass X-ray binaries \citep{Ertan2017,Ertan2018, Ertan2021}. 
In future work we will incorporate the realistic $\rin$ and torque calculations into our long-term evolution code to study evolutionary links between the different neutron star families in detail. Here we use our simpler model for a first comparison of the present and earlier results obtained with the same model. 

When accretion is allowed, we include the contribution of the cooling luminosity \citep{Page2006, Page2009} in the bolometric X-ray luminosity $\Lx = \Lacc + \Lcool$. The accretion luminosity $\Lacc~=~GM\Mdotstar/R_\ast$. Since $\rin$ is much greater than the radius of the star, $R_\ast$, contribution of the inner disc to $\Lx$ is negligible. In the SP phase, we take  $\Lx = \Lcool$. The X-rays produced by accretion and cooling processes are expected to be pulsed due to hotter polar regions of the star in both cases.
The cooling luminosity is important in the evolution of the sources, since it extends the lifetime of the disc even after $\Lacc$ becomes insufficient to keep the disc in the viscously active state. The outer disc radius, $\rout$, of the active disc is equal to the radius at which the current effective temperature is equal to the critical temperature $\Tp$. With decreasing X-ray irradiation strength, temperatures decrease throughout the disc, and $\rout$ moves inward. During the late phases of evolution, $\Lacc$ decreases sharply to below $\Lcool$, and subsequently the system enters the SP phase. Eventually, at ages close to $10^6$ yr, $\Lcool$ also decays sharply on entering the photon cooling era. The sharp drop in irradiation then quickly leads to inactivation of the entire disc.

\section{Results \& Discussion}
\label{results}

\begin{figure}
    \centering
    \includegraphics[width=1\columnwidth]{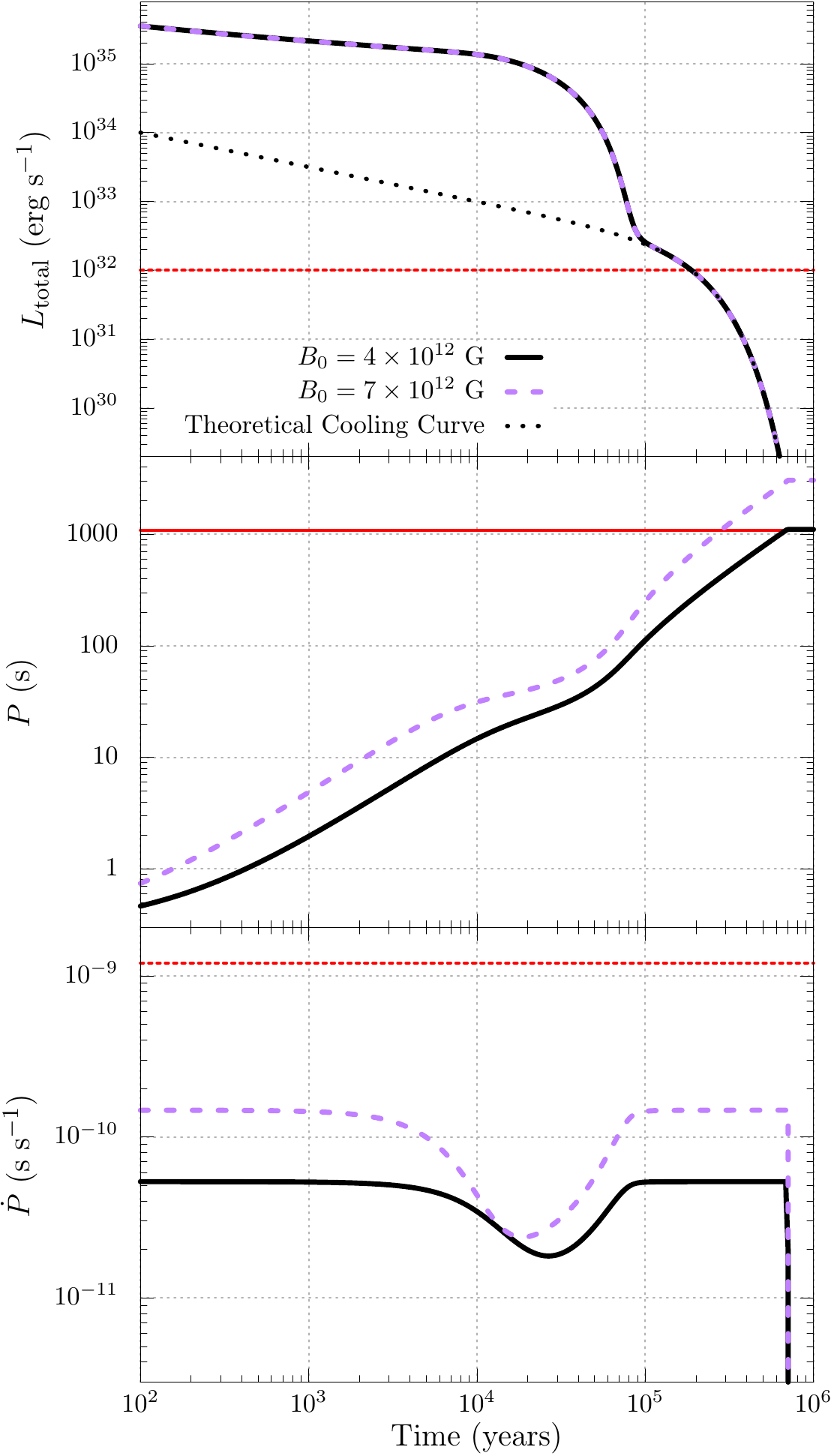}
    \caption{Illustrative model curves for the long-term evolution of \srca. $B_0$ values are given in the top panel. For both models, the other parameters are the same ($\alpha = 0.045$, $\Tp = 100$~K, $C = 7 \times 10^{-4}$, $P_0 = 0.3$~s, and $M_\mathrm{d} = 1.6 \times 10^{-5}~\Msun$). The horizontal dotted lines show the upper limits, $\Pdot < 1.2 \times 10^{-9}$~\spers~and $\Lx < 10^{32}$~\ergpers. In the top panel, the dotted curve shows the evolution of the cooling luminosity \citep{Page2006, Page2009}. The solid horizontal line in the middle panel shows the period of the source ($1091$~s).      
     }
    \label{fig1}
\end{figure}

The model curves in Fig. \ref{fig1} show illustrative evolutionary paths that reach the observed period of \srca~at a time when $\Lx$ and $\Pdot$ satisfy the observational upper limits (horizontal dotted lines). The model curves are obtained with disc parameters similar to those employed for other neutron star systems ($\alpha = 0.045$, $C = 7 \times 10^{-4}$, $\Tp = 100$~K), and with the initial conditions $P_0 = 0.3$~s, $\Md = 1.6 \times 10^{-5}~\Msun$, except $B_0$. The solid and dashed lines are for $B_0$ values of $4\times 10^{12}$~G and $7\times 10^{12}$~G respectively. The results are not sensitive to $P_0$ and $\Md$.

Viable models are obtained with $B_0 \simeq (4 - 8) \times 10^{12}$~G. For fields stronger than this range, the source attains the observed period at an earlier time when $\Lx$ is above the observational upper limit, while for fields below this range the disc becomes inactive before the observed period is reached. The model curves in Fig. \ref{fig1} are for $B_0 = 4 \times 10^{12}$~G (solid curves) and $B_0 = 7 \times 10^{12}$~G (dashed curves). The source is in the ASD phase at an age of $\sog 3 \times 10^5$~yr in both cases.The sharp drop in the $\Pdot$ curves takes place when the disc becomes completely inactive at time $\sim 7 \times 10^5$~yr when $P \sim 10^3$~s  for this source. Subsequently, the source evolves with the magnetic dipole torque alone, with 
$\Pdot \simeq 4 \times 10^{-18}$~\spers~for $B_0 = 7 \times 10^{12}$~G. For $B_0 \simeq 4 \times 10^{12}$~G, the source arrives at the $1091$~s period shortly before the inactivation of the disc and the period levels off at this value as the subsequent spin-down is negligible. This weaker field option is also a possibility for the current state of the source considering the large uncertainty in the $\Pdot$ measurement \citep{Hurley-Walker2022}. Further observations of the source during future radio burst epochs could better constrain the $\Pdot$ and the current state of the source.         

Long-term evolution of \srca~seems likely to be very similar to those of some of AXP/SGRs with relatively strong dipole moments. 
In earlier work using the same model, it was estimated that those AXP/SGRs could reach $10^3$~s periods 
\citep[see][fig. 2 and 3]{Benli2016}. Indeed, it is seen in Fig. \ref{fig1} that the long-$P$ pulsar would have been identified as an AXP/SGR if it were observed at earlier phases of evolution. The reason for the non-detection of such sources is likely to be their very faint X-ray luminosity and their lack of continuous, normal pulsed radio emission. Even in the case that the disc is active, it is not likely to be detected in the  infrared (IR) bands because of the weak X-ray irradiation and large $\rin$ ($\sim 2 \times 10^{10}$~cm). An evolutionary model for young neutron star families can be tested by observations of the progenitors of the long-period systems as proposed by that particular model. In our model, the AXP/SGRs are the progenitors expected to evolve to the long periods as exhibited by \srca. AXP/SGRs do have the high X-ray luminosities providing strong X-ray irradiation sufficient to irradiate and excite their discs to produce emission in the IR bands that should be detectable with the James Webb Space Telescope.

This long-$P$ pulsar was discovered by means of its radio burst events \citep{Hurley-Walker2022}. The observed pulsed radio emission obviously cannot be powered by magnetic dipole radiation alone, since the upper limit of $\Edot$ is smaller than the pulsed radio luminosity, while in normal pulsars the radio luminosity is orders of magnitude smaller than the rotational power.  At its present period, the inferred magnetic moment $\umu \sim 10^{30}$~G places \srca~below the radio pulsar death valley by four orders of magnitude. The physical mechanism producing the radio bursts is beyond the scope of the present work. Recently, \citet{Ronchi2022} proposed that a neutron star with a fallback disc and a magnetar dipole field can achieve the properties of this source at a young age($\sim 10^4 - 10^5$~yr). This model is completely different from the model we employ here. In their calculations, it is not clear why the disc is assumed to be inactive immediately after the source period is reached, while the current $\Mdotin$ is still high ($\sim 10^{15}$ \gpers). Furthermore, at these young ages $\Lcool$ alone is estimated to be well above the $\Lx$ upper limit, which is another source of heating for the disc that should be taken into account in the disc evolution models.

As a final remark, we would like to stress an important point about the comparison of the X-ray variability and the torque noise behavior of AXPs with the accreting neutron stars in binaries.  For this comparison two important criteria should be satisfied: (1) the neutron star should have a field strength and accretion rate  comparable to those estimated in the fallback disc model, and more importantly, (2) the dipole field of the neutron star in the binary should be interacting with a geometrically thin, optically thick disc, since AXPs in the fallback disc model have no companions. If the companion is a high-mass star, or a low-mass giant star, then their winds significantly affect both the X-ray variability and the torques acting on the neutron star. Analyses that do not take these criteria into account \citep[see e.g.][]{Doroshenko2020} produce misleading results.

\section{Conclusions}
\label{conc}

We have shown that the properties of \srca~can be accounted for as a product of long-term evolution in the fallback disc model. The long period of the source is obtained with a dipole moment of a few $10^{30}$ G cm$^3$ with the present state satisfying the upper bounds on $\Lx$ and $\Pdot$. Our results are not very sensitive to $\Md$ and $P_0$. There are two different evolutionary avenues compatible with the present source properties: (1) The disc is still active and the source is in the ASD phase at an age of $\sim 3 \times 10^5$~yr at present. For these solutions $P$ will continue to increase to a few $10^3$~s with $\Pdot \sim 10^{-10}$ \spers~until the disc becomes inactive at an age of $\sim 7 \times 10^5$~yr. (2) With a weaker field in the allowed range, the disc becomes inactive when the source achieves the observed period, and $P$ levels off at the present value because the dipole torque wields negligible spin-down after disc torques turn off: The source should be evolving with $\Pdot \sim 4 \times 10^{-18}$~\spers~at the present age greater than $\sim 7 \times 10^5$~yr. A tighter constraint on $\Pdot$ with future observations of the source could provide a test for these model predictions.  

\section*{Acknowledgements}

We acknowledge research support from Sabanc{\i} University, and from T\"{U}B\.{I}TAK (The Scientific and Technological Research Council of Turkey) through grant 120F329.

\section*{Data Availability}
No new data were analysed in support of this paper.
\bibliographystyle{mnras}
\bibliography{example} 








\bsp	
\label{lastpage}
\end{document}